\documentclass[12pt]{article}

\usepackage{amssymb,amsmath, amsfonts}
\usepackage{cases}
\usepackage{latexsym}
\usepackage{graphicx}
\usepackage{relsize}
\usepackage{natbib}  
\usepackage{rotating} 
\usepackage[includeheadfoot,left=1.in,right=1.in,top=1.in,bottom=1.in,bindingoffset=0.in,nohead]{geometry}
\usepackage[onehalfspacing]{setspace}
\usepackage{booktabs}  
\usepackage{appendix}
\usepackage[pdfborder={0 0 0}]{hyperref}
\usepackage{float}
\usepackage{xcolor}
\usepackage{caption} 
\usepackage{subcaption}
\usepackage{minibox}

\usepackage{tikz}
\usepackage{tabularx}
\newcolumntype{Y}{>{\centering\arraybackslash}X}

\usepackage[utf8]{inputenc}
\usepackage{subcaption}
\usepackage{url}
\usepackage[english]{babel}

\usepackage{mathptmx}
\usepackage[super]{nth}
\usepackage{graphicx, multicol}
\usepackage[flushleft]{threeparttable}
\usepackage{lscape}
\usepackage{multirow}
\usepackage{booktabs}
\usepackage{makecell}
\usepackage{enumitem}

\begin{document}

	\title{Optimal Combination of\\  Arctic Sea Ice Extent Measures:\\
A Dynamic Factor Modeling Approach}

	\author{Francis X. Diebold\\University of Pennsylvania \and Maximilian G\"obel\\ISEG - Universidade de Lisboa \and Philippe Goulet Coulombe\\University of Pennsylvania  \and Glenn D. Rudebusch\\Federal Reserve Bank of San Francisco \and Boyuan Zhang\\University of Pennsylvania\\$~$}

	\maketitle

	\begin{spacing}{1}

		 \noindent \textbf{Abstract}: The diminishing extent of Arctic sea ice is a key indicator of climate change as well as an accelerant for future global warming. Since 1978, Arctic sea ice has been measured using satellite-based microwave sensing; however, different measures of Arctic sea ice extent have been made available based on differing algorithmic transformations of the raw satellite data. We propose and estimate a dynamic factor model that combines four of these measures in an optimal way that accounts for their differing volatility and cross-correlations.  We then use the Kalman smoother to extract  an optimal combined measure of Arctic sea ice extent.  It turns out that almost all weight is put on the NSIDC Sea Ice Index, confirming and enhancing confidence in the Sea Ice Index and the NASA Team algorithm on which it is based.

		\thispagestyle{empty}

\smallskip

		\noindent {\bf Acknowledgments}: For comments and/or assistance we thank the editor, an associate editor, and two referees, as well as  Gladys Teng and the Penn Climate Econometrics Research Group. The views expressed here are those of the authors and do not necessarily represent those of others in the Federal Reserve System.

\smallskip


\smallskip


\smallskip

		\noindent {\bf Contact}: Philippe Goulet Coulombe,  gouletc@sas.upenn.edu

	\end{spacing}

\normalsize

	\clearpage

	\setcounter{page}{1}
	\thispagestyle{empty}

\section{Introduction}

Climate change is among the most pressing issues of our time, with many severe economic, environmental, and geopolitical consequences. Recently, the application of time series analytical methods to this topic -- and, more broadly, a ``climate econometrics" -- has emerged as a vibrant research literature, as highlighted, for example, in \cite{Hillebrandetal2020} and the references therein. One important issue that these methods can address is the loss of Arctic sea ice. The loss of Arctic sea ice is a vital focus point of climate study.  It is both an ongoing conspicuous \textit{effect} of climate change and a \textit{cause} of additional climate change via feedback loops.  In particular, reduced Arctic sea ice boosts solar energy absorption via decreased albedo due to darkening color (e.g., \cite{Stroeve2012}, \cite{PistoneEtAl2019},   \cite{DRice}) and increased methane release due to melting  permafrost  (e.g., \cite{VaksEtAl2020}).\footnote{For a broad and insightful overview of the evolution and causes of reduced Arctic sea ice cover,  see \cite{SeaIceBookCh4}.}

\begin{figure}[p]
	\caption{Four Sea Ice Extent Indicators}
	\begin{center}
		\includegraphics[scale=.3,trim={0 45mm 0 0},clip]{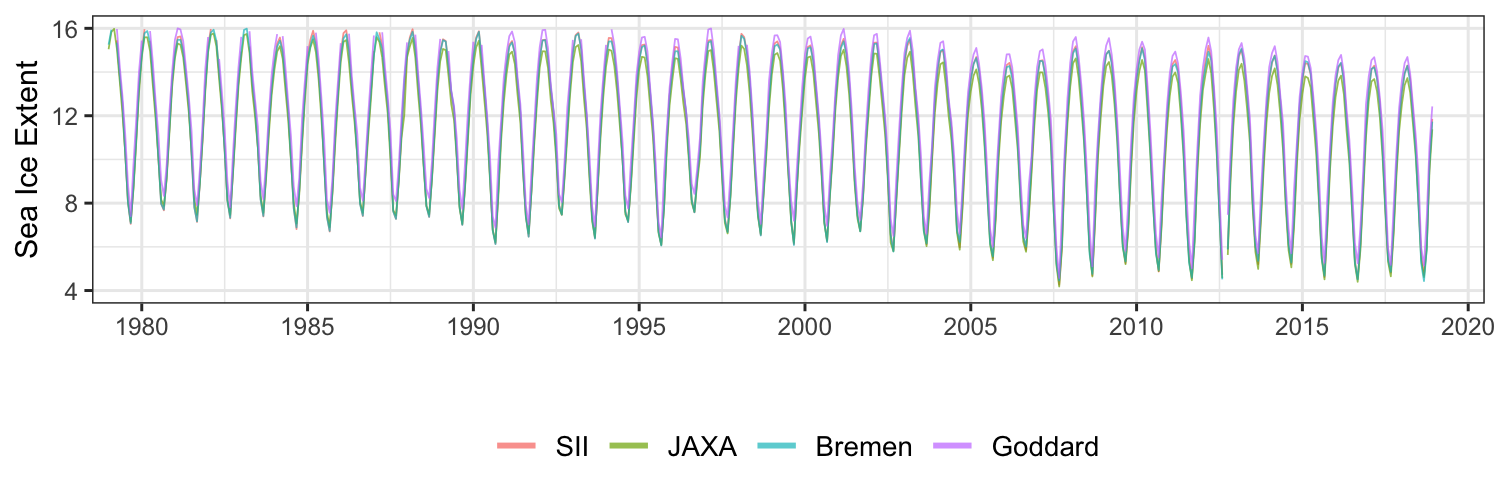}
		\includegraphics[scale=.25, trim={0 20 0 5mm},clip]{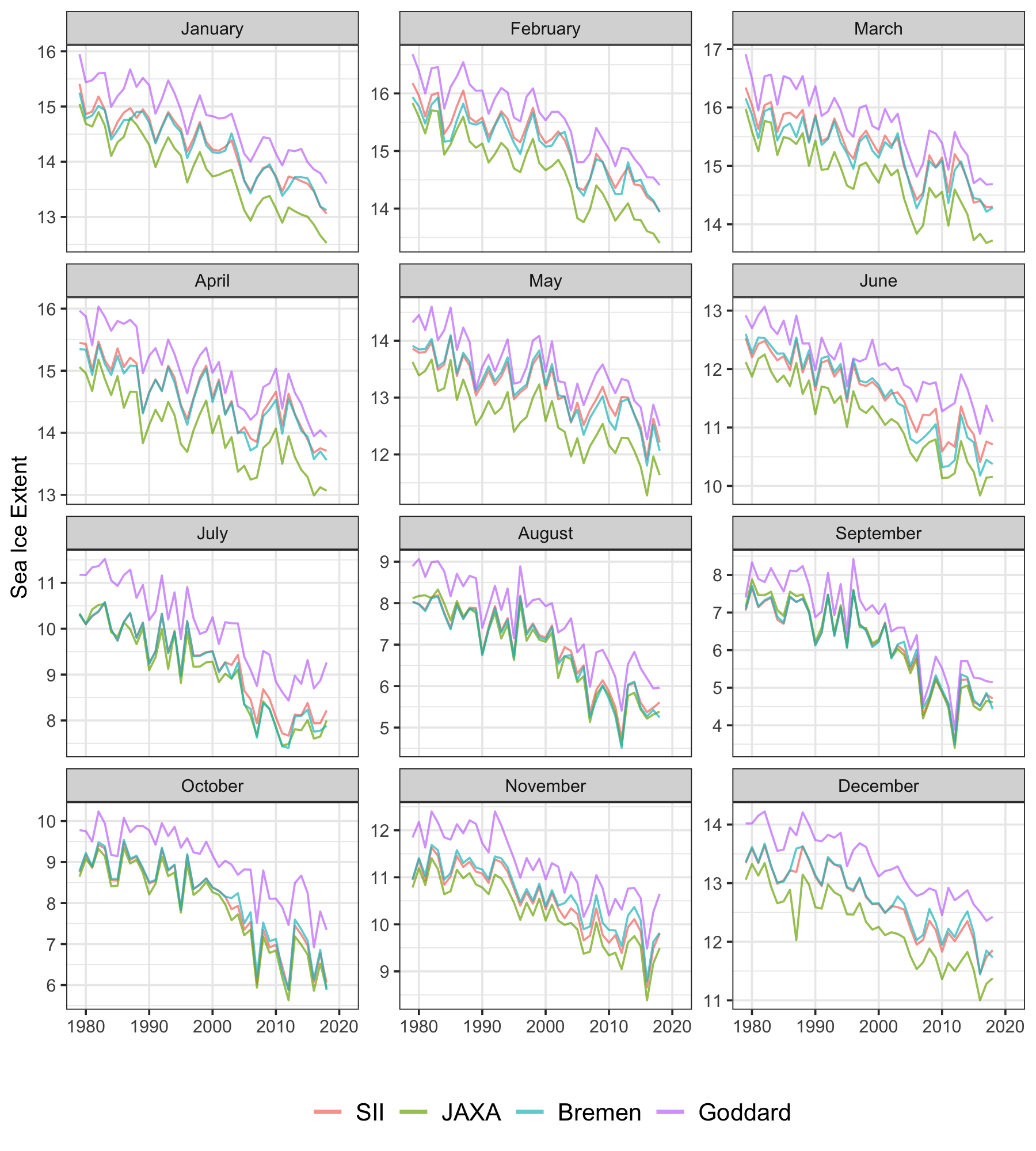}
	\end{center}
	\label{rawmonth1}
Notes: We show the Sea Ice Index (SII), Japan Aerospace Exploration agency (JAXA), University of Bremen (Bremen), and Goddard Bootstrap (Goddard). Units are millions of square kilometers.
\end{figure}

There are, however, several alternative measures of Arctic sea ice extent based on  different processing methodologies of the underlying satellite-based microwave measurement data, and the choice among these measures is not clear-cut (\cite{BunzelEtAl16}). In this paper, we study four such sea ice extent ($SIE$) measures, which we denote as Sea Ice Index ($SIE^S$), Goddard Bootstrap ($SIE^G$), JAXA ($SIE^J$), and Bremen ($SIE^B$). The top panel of Figure \ref{rawmonth1} provides time series plots of these four measures of Arctic $SIE$ for the satellite measurement era, which started in 1978. The four measures appear almost identical, because their scale is dominated by large seasonal swings. However, the effects of seasonality can be removed by plotting  each month separately for the four series, as done in the lower twelve panels of Figure \ref{rawmonth1}. Of course, the Arctic $SIE$ measures all trend down in every month, with steeper trends for the low-ice ``summer"  months (e.g., August, September, October). (Note the different axis scales for different months.) There are also systematic differences across indicators.   $SIE^G$, for example,  tends to be high, and  $SIE^J$ tends to be low, while $SIE^S$ and $SIE^B$ are intermediate.  But the deviations between various pairs of measures are not rigid; that is, they are not simply parallel translations of each other. Instead, there are sizable time-varying differences among the various measures. 

All of this suggests treating the various measures as noisy indicators of latent true sea ice extent, which in turn suggests the possibility of blending them into a single combined indicator with less measurement error.  Indeed some prominent studies have used simple equally-weighted averages of competing indicators, with precisely that goal. For example, a recent   report on the state of the cryosphere (\cite{IPCC19}) uses a simple average of three indicators.\footnote{See the notes for their Figure 3.3, page 3-13.}  Simple averages, however,  are  often sub-optimal. Optimality generally requires use of  \textit{weighted} averages giving, for example, less weight to noisier indicators. Motivated by these considerations, in this paper, we propose and explore a dynamic factor state-space model that combines the various published indicators into an optimal measure of sea ice extent, which we extract using the Kalman smoother.  

We proceed as follows. In section \ref{four}, we describe the four leading Arctic sea ice extent indicators that we study, and the satellites, sensors, and algorithms used to produce them.  In section \ref{extraction}, we propose a basic dynamic-factor state-space model for sea ice extent  and use it to obtain optimal extractions of latent extent.    We conclude in section \ref{concl}.

\section{Four Arctic Sea Ice Extent Indicators}  \label{four}

Sea ice extent ($SIE$) indicators are constructed from satellite measurements of the earth's surface using passive microwave sensing, which is unaffected by cloud cover or a lack of sunlight. Several steps are necessary to convert raw reflectivity observations into final $SIE$ measurements. First, for a polar region divided into a grid of individual cells, various sensors record a brightness reading or ``brightness temperature" for each cell. An algorithm then transforms these brightness readings into fractional surface coverage estimates -- sea ice concentration ($SIC$) values -- for each grid cell. Finally, $SIE$ is calculated by summing the area of all cells with at least 15 percent ice surface coverage.\footnote{\cite{PARKINSONETAL2008} discuss reasons for using a 15 percent cutoff.} This up-rounding in $SIE$ is effectively a bias correction, as determining the edge between ice and water can be especially difficult in the summer, when, for example, melting pools on summer ice surfaces can be mistaken for ice-free open water (\cite{MeierAndStewart2019}).

Different algorithms for processing the raw measurements importantly shape the final $SIE$ estimates. In addition, the $SIE$ series are not based on identical raw data because they use somewhat different satellites and sensors (\cite{ComisoEtAl17}, \cite{NSIDC_Boot_Explained}). In this section, we review some aspects of the satellites, sensors, and algorithms that underlie the $SIE$ measures. 

\subsection{Satellites and Sensors}

Table \ref{tab:Satellites} summarizes the operative dates of the various satellites and sensors relevant for Arctic sea ice measurement. The first multi-frequency sensor equipped on a satellite was the {Scanning Multichannel Microwave Radiometer} (SMMR)  launched in 1978 (\cite{NSIDC_Caval}). Starting in 1987, later sensors -- the Special Sensor Microwave Imager (SSM/I) and the {Special Sensor Microwave Imager/Sounder} (SSMIS) -- offered higher resolution images.\footnote{For detailed discussion of sensor characteristics see  \url{https://nsidc.org/ancillary-pages/smmr-ssmi-ssmis-sensors}.} In 2002 and 2012, respectively, the {Advanced Microwave Scanner Radiometer for EOS} (AMSR-E) and {Advanced Microwave Scanner Radiometer 2} (AMSR2) sensors were launched and provided further improvements in resolution (\cite{ComisoEtAl17}).\footnote{Early in the sample, operational problems prevented data delivery for several days during 1986 and between December 1987 and January 1988 (\cite{NSIDC_Boot}). For more recent technical difficulties, see \url{https://www.nrl.navy.mil/WindSat/Description.php}.}
Given the inclinations of the satellite orbits and the spherical shape of the earth, all of the satellites share an inability to observe the Arctic ``pole hole" -- a circular region at the very top of the world. The size of the pole hole varies across sensors, but historically, there is full confidence that the area covered by the pole hole fulfills the 15 percent  $SIC$ requirement (\cite{MeierAndStewart2019}). 

Table \ref{tab:Satellites} also describes the underlying source data for our four 
$SIE$ indicators. These measures use algorithms to transform the raw satellite brightness data into $SIC$ and $SIE$ values. We now turn to a more detailed discussion of these algorithms to illuminate the differences across $SIE$ indicators.

\begin{table} [tbp]
	\scriptsize
	\centering
	\caption{Satellites, Sensors, and Algorithms}\ \\
	\begin{threeparttable}
		\begin{tabularx}{\textwidth}{lYY|YY|YY|YY}
			\toprule 
						\multicolumn{1}{c}{\textbf{Satellite / Sensor}} & \multicolumn{2}{c}{\textbf{NASA Team}}  & \multicolumn{2}{c}{\textbf{Goddard Bootstrap}} & \multicolumn{2}{c}{\textbf{JAXA Bootstrap}} & \multicolumn{2}{c}{\textbf{ASI}} \\
			& \multicolumn{1}{c}{Start} & \multicolumn{1}{c}{End} & \multicolumn{1}{c}{Start} & \multicolumn{1}{c}{End} & \multicolumn{1}{c}{Start} & \multicolumn{1}{c}{End} & \multicolumn{1}{c}{Start} & \multicolumn{1}{c}{End} \\
			\cmidrule(lr){2-2} \cmidrule(lr){3-3} \cmidrule(lr){4-4} \cmidrule(lr){5-5} \cmidrule(lr){6-6} \cmidrule(lr){7-7} \cmidrule(lr){8-8} \cmidrule(lr){9-9}
			Nimbus-7 \ SMMR & 10/26/1978 & 08/20/1987 & 11/01/1978 & 07/31/1987 & 11/1978 & 07/1987 & 1972 & ... \\
			DMSP-F8  \ SSM/I & 08/21/1987 & 12/18/1991 & 08/01/1987 & 12/17/1991 & 07/1987 & ... & ... & ...\\
			DMSP-F11  \ SSM/I & 12/19/1991 & 09/29/1995 & 12/18/1991 & 05/09/1995 & ... & ... & ... & ...\\
			DMSP-F13  \ SSM/I & 09/30/1995 & 12/31/2007 & 05/10/1995 & 12/31/2007 & ... & 06/2002 & ... & 12/2010\\
			DMSP-F17  \ SSMIS & 01/01/2008 & 12/31/2017 & 01/01/2008 & present & & & 10/2011 & 07/2012\\
			DMSP-F18  \ SSMIS & 01/01/2018 & present &  &  & & &  & \\
			EOS/Aqua  \ AMSR-E & & & & & 06/2002 & 10/2011 & 2003 & 10/2011 \\
			Coriolis  \ WindSat & & & & & 10/2011 & 07/2012 & &\\
			GCOM-W1  \ AMSR2 & & & & & 07/2012 & present & 07/2012 & present  \\
			\bottomrule 
		\end{tabularx}%
\footnotesize 
$~$\\
		Notes: NASA Team and Goddard Bootstrap dates from \cite{NSIDC_SII}. JAXA Bootstrap dates from \url{https://kuroshio.eorc.jaxa.jp/JASMES/climate/index.html}. ASI dates from \url{https://seaice.uni-bremen.de/sea-ice-concentration/time-series/}.
	\end{threeparttable}
	\label{tab:Satellites}%
\end{table}

\subsection{From Raw Measurement to $SIE$: Algorithmic Transformations}

Once brightness data have been recorded by satellite sensors, an algorithm converts the measurements into estimates of $SIC$. Here, we discuss the algorithms and other details of the various $SIE$ indicators.

\subsubsection{Sea Ice Index} \label{SII}

Updated on a daily basis and distributed by the {National Snow and Ice Data Center} (NSIDC), the Sea Ice Index (SII or $SIE^S$) 
 combines two separate Sea Ice indicators: (1) the {Sea Ice Concentrations from Nimbus-7 SMMR and DMSP SSM/I-SSMIS Passive Microwave Data} (NASA Team) (\cite{NSIDC_Caval}) -- produced at the {Goddard Space Flight Center} -- and (2) the {Near-Real-Time DMSP SSMIS Daily Polar Gridded Sea Ice Concentrations} (NRTSI) (\cite{NSIDC_NearReal}) -- produced by the NSIDC itself.\footnote{See   \url{https://doi.org/10.7265/N5K072F8}.}  A time-lag of about one year between the $SIE$ estimates by NASA Team and its publication in the NSIDC database requires the NRTSI to complement the SII.
 
The NRTSI follows the NASA Team algorithm as closely as possible, but inconsistencies between the two series cannot be ruled out entirely (\cite{NSIDC_SII}). 
In particular, the two sub-indicators use brightness temperatures from different providers.\footnote{\cite{NSIDC_NearReal} takes the data from the National Oceanic and Atmospheric Administration Comprehensive Large Array-data Stewardship System (NOAA CLASS); (\cite{NSIDC_Caval}) uses data processed at the NASA Goddard Space Flight Center.} These raw readings can be distorted by weather effects, making open water look like sea ice cover. Therefore, post-calculation quality checks apply land and ocean masks, to remove errorneous and implausible ice covers. However, NASA Team and NRTSI do not apply the exact same filters (\cite{NSIDC_SII}). The former algorithm additionally screens the data manually for falsely detected ice formation (\cite{NSIDC_Caval}), which can enhance accuracy but also reduce transparency of the final measurements.

As Table \ref{tab:Satellites} shows, the SII obtains the raw data from different generations of satellites and sensors.
To make the data comparable, a linear least-squares model on the brightness temperatures, as reported by the two distinct sensors for an overlapping period of operation, is intended to adjust the reference points of 100 percent sea ice and 100 percent open water. These \emph{tie points} then remain fixed over the lifetime of the new system (\cite{NSIDC_Caval_tie}).

\subsubsection{Goddard Bootstrap} \label{Bootstrap}

Another sea ice indicator, distributed by the NSIDC, relies on $SIC$ estimates from the Goddard Bootstrap algorithm ($SIE^G$).\footnote{For detailed algorithm description see \cite{ComisoEtAl17b}. For the data, see \url{https://doi.org/10.5067/7Q8HCCWS4I0R}.}  Despite the NASA Team and the Goddard Bootstrap algorithms having both been developed at the \emph{NASA Goddard Space Flight Center}, there are some differences between the two approaches.
These arise mostly from the calibration of tie points: While the NASA Team adjusts these reference points for 100 percent open water and 100 percent ice only when a new satellite or sensor becomes operational, the Goddard Bootstrap algorithm adjusts these reference points on a daily basis to account for varying weather conditions (\cite{ComisoEtAl17}).
Differing weather filters and sensitivities to varying physical temperature also lead to differences in the final measurements(\cite{ComisoEtAl97}). In contrast to the NASA Team, the strength of the Goddard Bootstrap algorithm is the identification of melting sea ice. Therefore, the Goddard Bootstrap algorithm provides more accurate estimates of the edge of the ice cover (\cite{goldsteinetal2018}).

Although differences between $SIE^S$ and $SIE^G$ are generally assessed to be small, they cannot necessarily be neglected (\cite{goldsteinetal2018}).\footnote{See also  \url{https://nsidc.org/support/faq/nasa-team-vs-bootstrap-algorithm}.} The differences between the NASA Team and Goddard Bootstrap algorithms occur especially during the melting season, when the former generally reports larger deviations from ship or radar observations. However, the relative accuracy of the two algorithms is not clear-cut, as the Goddard Bootstrap algorithm is highly sensitive to physical temperature and underestimates $SIC$ during winter times in the higher latitudes of the Arctic region.

\subsubsection{Japan Aerospace Exploration Agency}

As listed in Table \ref{tab:Satellites}, both $SIE^S$ and $SIE^G$ rely on the same set of instruments, which have been criticized for their low spatial resolution (\cite{goldsteinetal2018}). The Japan Aerospace Exploration Agency (JAXA) sea ice measure ($SIE^J$) uses an adapted version of the Goddard Bootstrap algorithm to derive $SIC$ measures from satellite readings with higher spatial resolution (\cite{ComisoEtAl17b}).\footnote{For description and data see \url{https://kuroshio.eorc.jaxa.jp/JASMES/climate/index.html}.} However, readings from these high resolution satellites are only available since 2000, so their data must merged with observations from older sensors to extend the data coverage to 1978 (\cite{ComisoEtAl17}). $SIE^J$ also distinguishes itself from $SIE^S$ and $SIE^G$ by using a 5-day moving average of observations to compensate for potentially missing data.

\subsubsection{University of Bremen}

Using observations delivered by the high-resolution AMSR-E sensor, a group of researchers at the University of Bremen developed the ARTIST Sea Ice (ASI) algorithm (\cite{SpreenEtAl08}) to estimate  daily $SIC$.\footnote{Monthly data  from  \url{https://seaice.uni-bremen.de/data/amsr2/today/extent_n_19720101-20181231_amsr2.txt}.} The time-series ($SIE^B$) uses different algorithms for different sensors. Until the launch of the AMSR-E sensor in 2003, $SIE^B$ used the NASA Team algorithm to transform brightness readings into $SIC$ values. From then on, the NASA Team algorithm is replaced by the ASI.\footnote{See  \url{https://seaice.uni-bremen.de/sea-ice-concentration-amsr-eamsr2/time-series/}.}

\section{Optimal Extraction of Latent Extent} \label{extraction}

The four sea ice indicators  discussed above differ in terms of the raw data sources and the algorithms used to process the raw  data.  They can be viewed as distinct indicators of an unobserved or latent ``true'' sea ice extent, $SIE^*$.  Blending such a set of noisy indicators can produce a single series with less measurement error.  Here we formalize this intuition in a state-space dynamic-factor model, from which we extract an optimal composite estimate of  $SIE^*$ from the four component indicators.\footnote{Dynamic factor analysis is closely related to principal components analysis, but the dynamic factor model provides a fully-specified probabilistic modeling framework, in which  estimation and factor extraction via the Kalman filter are statistically efficient.  Under conditions the two approaches coincide in large samples, but those conditions include a very large number of indicators.  Those conditions  are violated in our case as we have only four indicators, so dynamic factor analysis is preferable.  For a much more complete account, see \cite{SW2011}.}

\subsection{A Dynamic Factor Model} \label{combining}

We work in a state-space environment, modeling each of the four indicators ($SIE^S$, $SIE^J$, $SIE^B$, and $SIE^G$) as driven by latent true sea ice extent, $SIE^*$, with an additive measurement error.\footnote{The approach parallels \cite{ADNSS2016}, who extract latent ``true" U.S. GDP from noisy expenditure-side and income-side estimates.}
As discussed previously, some indicators present level shifts with respect to one another, most often resulting from how they respectively deal with tie points. It is thus preferable to constrain both the trend and dynamics to follow a common factor, while leaving level offsets unconstrained. The measurement equation is

\begin{equation} \label{meas99}
\begin{pmatrix}
SIE^S_{t} \\ SIE^J_{t} \\ SIE^B_{t}  \\ SIE^G_{t}
\end{pmatrix}
 =
 \begin{pmatrix}
	c_S \\ c_J \\ c_B \\ c_G
 \end{pmatrix}
 +
\begin{pmatrix}
\lambda_S \\ \lambda_J \\ \lambda_B \\ \lambda_G
\end{pmatrix}
SIE^*_t
+
\begin{pmatrix}
\varepsilon^S_{t} \\ \varepsilon^J_{t} \\ \varepsilon^B_{t} \\ \varepsilon^G_{t}
\end{pmatrix},
\end{equation}
where
\begin{equation} \label{stoc}
\varepsilon_t = (\varepsilon^S_{t},  \varepsilon^J_{t},  \varepsilon^B_{t},  \varepsilon^G_{t})'
~ \sim ~ iid   \left (0 , ~ \Sigma \right ),
\end{equation}
with
\begin{equation} \label{covmat}
 \Sigma =
\begin{pmatrix}
\sigma^2_{SS} & \cdot  &  \cdot & \cdot  \\
\sigma^2_{JS} &  \sigma^2_{JJ} & \cdot & \cdot  \\
\sigma^2_{BS} &  \sigma^2_{BJ} &  \sigma^2_{BB} & \cdot  \\
\sigma^2_{GS} &  \sigma^2_{GJ} &  \sigma^2_{GB} & \sigma^2_{GG}  \\
\end{pmatrix}.
\end{equation}
Moreover, $c_i = 0$ when we normalize $\lambda_i = 1$ for $i \in \{S, J, B, G\}$. 

The transition equation is
\begin{equation} \label{trans}
	SIE^*_t  = \rho \, SIE^*_{t-1} + TREND_t +  SEASONAL_t +  \eta_t,
\end{equation}
where $\eta_t \sim iid (0, \sigma^2_{\eta \eta})$ is orthogonal to $\varepsilon_t$ at all leads and lags.  Various modeling approaches are distinguished by their treatment of $TREND_t$ and $SEASONAL_t$.   We follow \cite{DRice} and allow for 12 monthly deterministic seasonal effects, each of which is endowed with (possible) deterministic quadratic trend.\footnote{We emphasize that our  model is meant to be a simple benchmark, and that many potentially-important variations and extensions are possible.  For example,  one could alternatively  entertain stochastic as opposed to deterministic trend and  seasonality.  A simple approach would be separate month-by month modeling  so that there is no seasonality, whether with one unit root, as in (for month $m$) $	TREND_{m,t} = d_m + TREND_{m,t-1} + u_{m,t}$, or two unit roots, as in $TREND_{m,t} = d_{mt} + TREND_{m,t-1} + u_{m,t}$, where $d_{m,t} = d_{m,t-1} + v_{m,t}$.
}
  This results in a blended deterministic ``trend/seasonal" given by
\begin{equation}
	TREND_t +  SEASONAL_t = \sum_{i=1}^{12} a_i \, D_{it} + \sum_{j=1}^{12} b_j \,  D_{jt} {\cdot} TIME_t + \sum_{k=1}^{12} c_k \,  D_{kt} {\cdot} TIME_t^2,
\end{equation}
where  $D_i$ indicates month $i$ and $TIME$ indicates time. Hence the full transition equation is

\begin{equation} 	\label{full99}
	SIE^*_t = \rho SIE^*_{t-1} + \sum_{i=1}^{12} a_i  \, D_{it} + \sum_{j=1}^{12} b_j  \, D_{jt} {\cdot} TIME_t + \sum_{k=1}^{12} c_k  \, D_{kt} {\cdot} TIME_t^2  + \eta_t.
\end{equation}

The model is already in state-space form, and one pass of the Kalman filter, initialized with the unconditional state mean and covariance matrix,  provides the 1-step prediction errors necessary to construct the Gaussian pseudo-likelihood, which we maximize using the EM algorithm and calculate standard errors from the analytic Hessian matrix.\footnote{Note that we do not assume Gaussian shocks, and that it is not necessary to assume Gaussian shocks, as we can still maximize the Gaussian likelihood even if the shocks are not truly Gaussian, and the resulting (pseudo-)MLE still has the good properties of consistency,  asymptotic normality, etc.} Following estimation, we use the Kalman smoother to obtain the best linear unbiased extraction of $SIE^*$ from the estimated model. The smoother averages across indicators, but it desirably produces optimally weighted averages rather than simple averages.  The smoother also averages over time, using data both before and after time $t$ to estimate $SIE_t^*$, which is also necessary for optimal extraction, due to the serial correlation in $SIE^*$. For details see \cite{Harvey1989}.

One or more  restrictions are  necessary for identification.  The standard approach is to normalize a factor loading, which amounts to an unbiasedness assumption.  Normalizing   $\lambda_S {=} 1$, for example, amounts to an assumption that $SIE^S$ is unbiased for  $SIE^*$.  Whether there truly exists such an unbiased indicator (and if so, which) is of course an open question -- one can never know for sure.   $SIE^S$ and  $SIE^G$ are the most widely used indicators (\cite{MeierEtAl14},  \cite{PengEtAl13}), so it is natural to consider normalizing on  $\lambda_S$ or  $\lambda_G$.  We explore both.

\subsection{Estimated Measurement Equation}

The estimated  measurement equation (\ref{meas99}), normalized with $\lambda_S  {=}  1$, is 

\begin{equation}  \label{222}
\begin{pmatrix}
	SIE^S_t \\ SIE^J_t \\ SIE^B_t  \\ SIE^G_t
\end{pmatrix}
=
\begin{pmatrix}
0\\ \underset{[0.025]}{0.225} \\ \underset{[0.020]}{0.043} \\ \underset{[0.034]}{1.040}
\end{pmatrix}
+
\begin{pmatrix}
	1 \\ \underset{[0.002]}{0.950} \\ \underset{[0.002]}{0.995} \\ \underset{[0.003]}{0.961}
\end{pmatrix}
SIE^*_t
+
\begin{pmatrix}
	\varepsilon^S_{t} \\ \varepsilon^J_{t} \\ \varepsilon^B_{t} \\ \varepsilon^G_{t}
\end{pmatrix} ,
\end{equation}
where standard errors appear beneath each estimated loading.   All indicators are estimated to load  heavily on $SIE^*$, with all $\hat{\lambda}'s$ very close to 1.   $SIE^J$ and $SIE^G$ load least heavily ($\hat{\lambda_J}  {=}  0.950$, $\hat{\lambda_G} {=}  0.961$)), in accord with their generally less abrupt trend in Figure \ref{rawmonth1}.  $SIE^B$ loads with an estimated coefficient marginally different from 1 but significant at the 5\% level. Of course, the $SIE^S$ loading is 1 by construction. While Bremen's level offset (with respect to $SII$) is arguably negligible, those of Jaxa and especially Goddard are sizable.  Hence, the estimation results accord with Figure \ref{rawmonth1}, with  $SIE^S$ and  $SIE^B$ more in the center of the range and  $SIE^J$ and  $SIE^G$ more extreme.

Alternatively, the estimated  measurement equation normalized with $\lambda_G  {=}  1$ is 

 \begin{equation}   \label{222a}
 \begin{pmatrix}
 SIE^S_t \\ SIE^J_t \\ SIE^B_t  \\ SIE^G_t
 \end{pmatrix}
 =
 \begin{pmatrix}
\underset{[0.039]}{-1.081} \\ \underset{[0.038]}{-0.803} \\ \underset{[0.042]}{-1.033} \\ 0
 \end{pmatrix}
 +
 \begin{pmatrix}
	\underset{[0.003]}{1.040}\\ \underset{[0.003]}{0.988} \\ \underset{[0.003]}{1.034} \\ 1
 \end{pmatrix}
 SIE^*_t
 + 
 \begin{pmatrix}
 \varepsilon^S_{t} \\ \varepsilon^J_{t} \\ \varepsilon^B_{t} \\ \varepsilon^G_{t}
 \end{pmatrix}.
 \end{equation}
 The estimated loadings in equations (\ref{222}) and (\ref{222a}), corresponding to $\lambda_S  {=}  1$ and $\lambda_G  {=}  1$ respectively, are effectively identical up to the normalization.  

Now consider the associated measurement error covariance matrix (\ref{covmat}).  The estimate for the $\lambda_S  {=}  1$ normalization is

\begin{equation}  
\hat{\Sigma} =
\begin{pmatrix}
\underset{[0.0042]}{0.0003} &  \cdot  &  \cdot  &  \cdot   \\
\underset{[0.0043]}{0.0010} & \underset{[0.0049]}{0.0236} &  \cdot  &   \cdot  \\
\underset{[0.0044]}{0.0004} & \underset{[0.0045]}{0.0025} & \underset{[0.0048]}{0.0146} & \cdot  \\
\underset{[0.0044]}{-0.0025} & \underset{[0.0048]}{0.0081} & \underset{[0.0046]}{0.0002} & \underset{[0.0056]}{0.0361}  \\
\end{pmatrix},
\end{equation}
with implied estimated correlation matrix 
\begin{equation}  
\hat{R} = 
\begin{pmatrix}
1 & \cdot & \cdot  &  \cdot \\
0.366 & 1  & \cdot  &  \cdot \\
0.206 & 0.134  & 1 & \cdot \\
-0.747 & 0.276 & 0.008  & 1 \\
\end{pmatrix}.
\end{equation}
Note that $\hat{\sigma}^2_{GG}$ is much higher than any of $\hat{\sigma}^2_{SS}$, $\hat{\sigma}^2_{JJ}$, and $\hat{\sigma}^2_{BB}$, potentially due to different indicators using different methods to determine tie points, i.e., reference points of brightness for 100\% sea ice and 100\% open water. The choice is crucial for accurate measurement of $SIC$ within grid cells.  Tie points, moreover, need not be constant, as brightness readings are sensitive to weather effects and atmospheric forcings (\cite{Ivan15}). Dynamic tie-point calibration is potentially desirable because it can decrease the bias of $SIC$ measurements (\cite{ComisoEtAl17}). The latest version of the Goddard Bootstrap algorithm, in particular, calibrates  tie points daily. One would expect, however, that the bias reduction from dynamic tie-point calibration may come at the cost of potential discontinuities that increase measurement error variance. Our results confirm that conjecture.  Our estimate of $\hat{\sigma}^2_{GG}$ is about 13 times that of $\hat{\sigma}^2_{BB}$ (which uses constant tie points).

Alternatively, the estimated measurement error covariance matrix for the  $\lambda_G  {=}  1$ normalization is 

\begin{equation}  
\hat{\Sigma} =
\begin{pmatrix}
\underset{[0.0041]}{0.0003} &  \cdot  &  \cdot  &  \cdot   \\
\underset{[0.0041]}{0.0008} & \underset{[0.0048]}{0.0233} &  \cdot  &   \cdot  \\
\underset{[0.0042]}{0.0003} & \underset{[0.0044]}{0.0022} & \underset{[0.0047]}{0.0144} & \cdot  \\
\underset{[0.0042]}{-0.0025} & \underset{[0.0046]}{0.0079} & \underset{[0.0045]}{0.0001} & \underset{[0.0055]}{0.0361}  \\
\end{pmatrix},
\end{equation}
with implied estimated correlation matrix 
\begin{equation}
\hat{R} = 
\begin{pmatrix}
1 & \cdot & \cdot  &  \cdot \\
0.311 & 1 &\cdot  & \cdot  \\ 
0.161 & 0.121 & 1 & \cdot  \\ 
-0.782 & 0.272 & 0.002 & 1 \\ 
\end{pmatrix}.
\end{equation}

\subsection{Estimated Transition Equation}

\begin{table}[tbp]
	\begin{center}
		\caption{Trend/Seasonal Parameter Estimates}
		\label{tab:trans_eq} 
		\begin{tabular}{c r r r r@{\hskip 0.4in} r r r}
			\toprule
			& \multicolumn{3}{c}{$\lambda_S  {=}  1$} & & \multicolumn{3}{c}{$\lambda_G  {=}  1$}\\
			\cline{2-8} \noalign{\smallskip}
			& \multicolumn{1}{c}{$a_i$ } & \multicolumn{1}{c}{$b_j$ } & \multicolumn{1}{c}{$c_k$ }   & & \multicolumn{1}{c}{$a_i$ } & \multicolumn{1}{c}{$b_j$ } & \multicolumn{1}{c}{$c_k$ } \\ 
			\midrule
			\vspace{2mm}
			Jan &  $\underset{[0.583]}{5.287}$ &  $\underset{[1.005]}{1.412}$ &  $-\underset{[2.025]}{4.736}$ & & 
			$\underset{[0.602]}{5.245}$ &  $\underset{[0.953]}{0.876}$ &  $-\underset{[1.930]}{3.608}$ \\ 
			\vspace{2mm}
			Feb &  $\underset{[0.632]}{5.296}$ &  $-\underset{[0.995]}{0.31}$ &  $-\underset{[2.031]}{1.735}$ & &  $\underset{[0.653]}{5.171}$ &  $-\underset{[0.948]}{0.267}$ &  $-\underset{[1.936]}{1.616}$ \\ 
			\vspace{2mm}
			Mar &  $\underset{[0.666]}{4.858}$ &  $-\underset{[0.997]}{1.227}$ &  $\underset{[2.033]}{0.99}$ & &  $\underset{[0.686]}{4.738}$ &  $-\underset{[0.950]}{1.153}$ &  $\underset{[1.938]}{1.011}$ \\ 
			\vspace{2mm}
			Apr &  $\underset{[0.672]}{4.066}$ &  $-\underset{[1.003]}{1.719}$ &  $\underset{[2.027]}{2.187}$  & &  $\underset{[0.692]}{3.975}$ &  $-\underset{[0.956]}{1.62}$ &  $\underset{[1.932]}{2.133}$ \\ 
			\vspace{2mm}
			May &  $\underset{[0.644]}{2.977}$ &  $\underset{[1.012]}{0.367}$ &  $-\underset{[2.026]}{2.25}$ & &  $\underset{[0.665]}{2.939}$ &  $\underset{[0.965]}{0.398}$ &  $-\underset{[1.931]}{2.176}$ \\ 
			\vspace{2mm}
			Jun &  $\underset{[0.580]}{2.732}$ &  $-\underset{[1.011]}{1.307}$ &  $-\underset{[2.029]}{1.072}$ & &  $\underset{[0.603]}{2.725}$ &  $-\underset{[0.963]}{1.225}$ &  $-\underset{[1.934]}{1.021}$ \\ 
			\vspace{2mm}
			Jul &  $\underset{[0.525]}{1.657}$ &  $-\underset{[1.015]}{1.497}$ &  $-\underset{[2.031]}{3.053}$ & &  $\underset{[0.550]}{1.711}$ &  $-\underset{[0.967]}{1.411}$ &  $-\underset{[1.935]}{2.884}$ \\ 
			\vspace{2mm}
			Aug &  $\underset{[0.444]}{0.719}$ &  $\underset{[1.029]}{0.511}$ &  $-\underset{[2.041]}{5.634}$ & &  $\underset{[0.472]}{0.837}$ &  $\underset{[0.981]}{0.532}$ &  $-\underset{[1.945]}{5.338}$ \\ 
			\vspace{2mm}
			Sep &  $\underset{[0.350]}{1.715}$ &  $-\underset{[1.026]}{1.066}$ &  $-\underset{[2.064]}{2.894}$ & &  $\underset{[0.380]}{1.829}$ &  $-\underset{[0.978]}{0.993}$ &  $-\underset{[1.968]}{2.667}$ \\ 
			\vspace{2mm}
			Oct &  $\underset{[0.324]}{3.923}$ &  $\underset{[1.026]}{1.274}$ &  $-\underset{[2.061]}{5.929}$ & &  $\underset{[0.354]}{3.958}$ &  $\underset{[0.978]}{1.264}$ &  $-\underset{[1.965]}{5.583}$ \\ 
			\vspace{2mm}
			Nov &  $\underset{[0.393]}{4.976}$ &  $-\underset{[1.025]}{1.711}$ &  $\underset{[2.089]}{3.678}$ & &  $\underset{[0.421]}{4.947}$ &  $-\underset{[0.976]}{1.639}$ &  $\underset{[1.992]}{3.715}$ \\ 
			\vspace{2mm}
			Dec &  $\underset{[0.484]}{5.504}$ &  $-\underset{[1.032]}{0.791}$ &  $\underset{[2.036]}{0.056}$ & &  $\underset{[0.510]}{5.423}$ &  $-\underset{[0.984]}{0.738}$ &  $\underset{[1.941]}{0.133}$ \\ 
			\bottomrule
		\end{tabular}
	\end{center}
	
	\noindent Notes: The $b_j$ are $\times 10^{3}$ and the $c_k$ are $\times 10^{6}$.
\end{table}

Now let us move to the transition equation (\ref{full99}).  Using the $\lambda_S  {=}  1$ normalization we obtain $\hat{\rho} {=} 0.704 ~ [0.041]$ and  trend/seasonal parameter estimates ($\hat{a}_i$, $\hat{b}_j$, and $\hat{c}_k$) as reported in the $\lambda_S {=} 1$ columns of Table \ref{tab:trans_eq}. Trends for all months are highly significant and downward sloping.  The trends for summer months (August-November) display a notable negative, and generally statistically significant, {curvature}, whereas for non-summer months the quadratic trend terms are generally small and statistically  insignificant. For the $\lambda_G  {=}  1$ normalization we get $\hat{\rho} {=} 0.719~ [0.042]$ and  trend/seasonal parameter estimates ($\hat{a}_i$, $\hat{b}_j$, and $\hat{c}_k$) as reported in the  $\lambda_G {=} 1$ columns of Table \ref{tab:trans_eq}.  The $\lambda_S {=} 1$ and $\lambda_G {=} 1$  results are very similar.

\subsection{Extracted Latent Sea Ice Extent}

\begin{figure}[tbp]
	\caption{Extracted  Sea Ice Extent and Four Raw Indicators, by Month, $\lambda_S {=}  1$}
	\begin{center}
		\includegraphics[scale=.25]{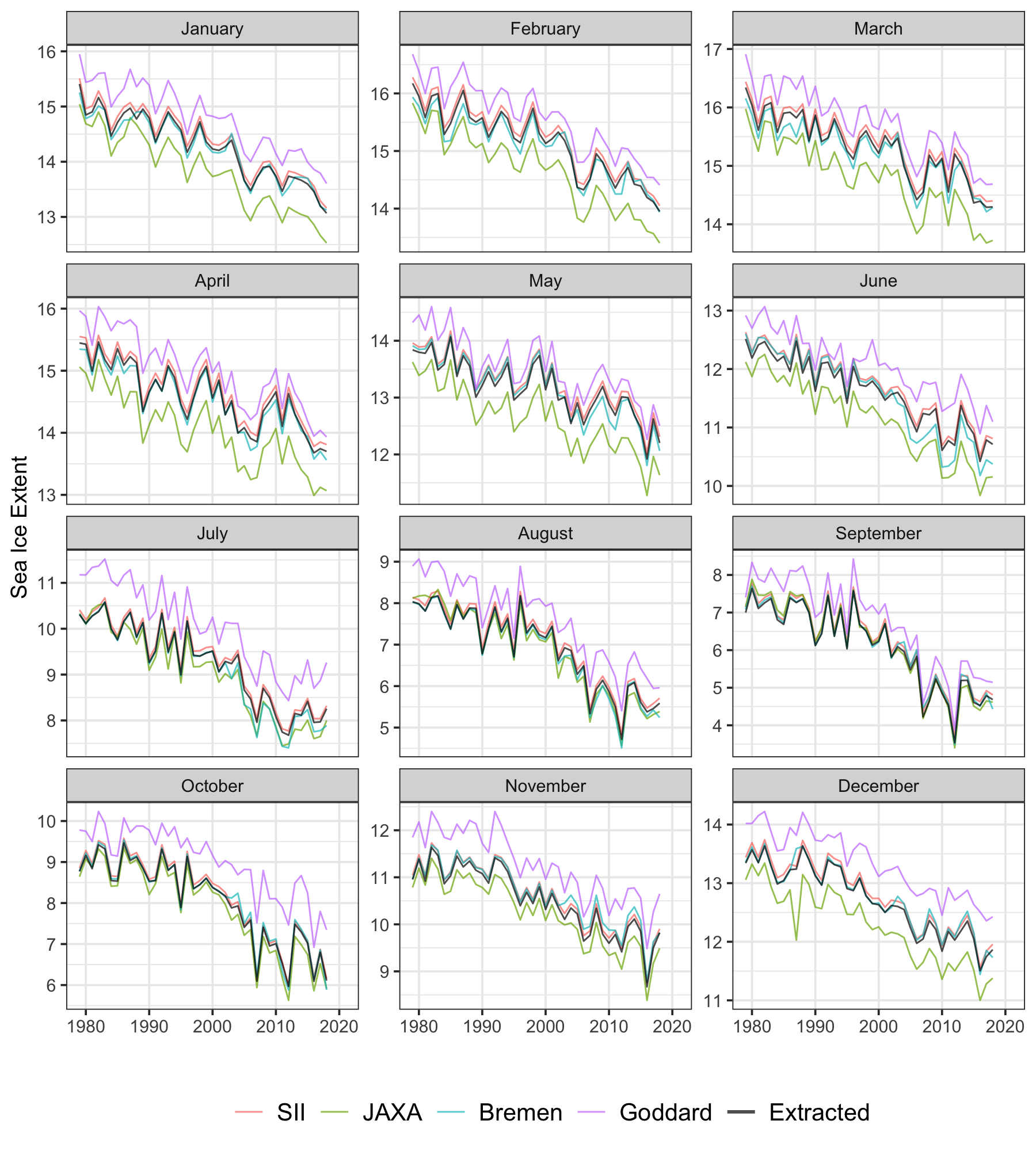}
	\end{center}
	\label{ice_dts_month}
	Notes: We show sea ice extent extracted assuming $\lambda_S {=} 1$, together with four raw indicators: Sea Ice Index (SII), Japan Aerospace Exploration agency (JAXA), University of Bremen (Bremen), and Goddard Bootstrap (Goddard). Units are millions of square kilometers.  
\end{figure}

\begin{figure}[tbp] 
	\caption{Extracted Latent Sea Ice Extents and Four Raw Indicators, by Month, $\lambda_G  {=}  1$}
	\begin{center}
		\includegraphics[scale=.25]{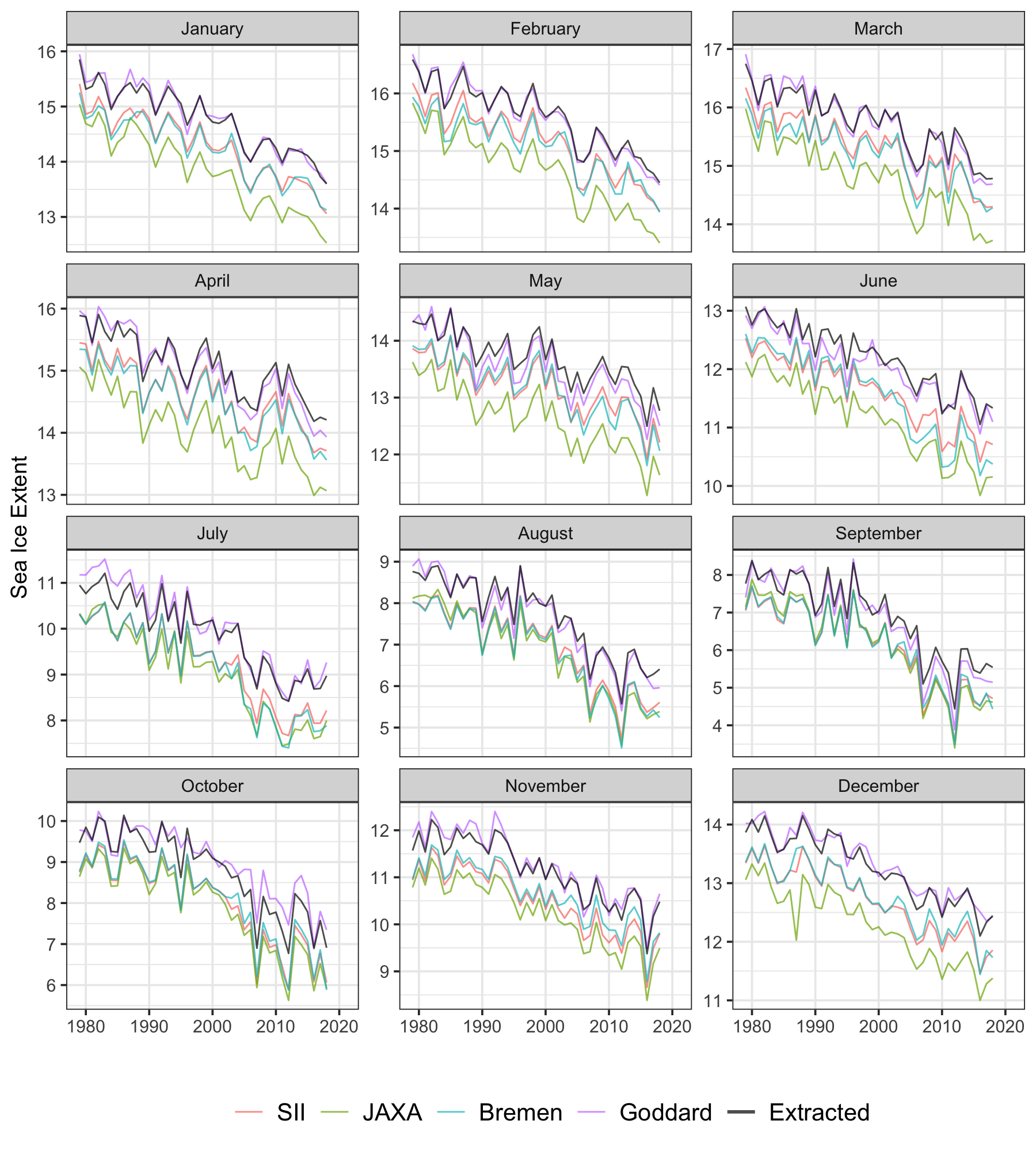}
	\end{center}
	\label{ice_dts_month_seqnseq}
	Notes: We show sea ice extent extracted assuming $\lambda_G {=} 1$, together with four raw indicators:  Sea Ice Index (SII), Japan Aerospace Exploration agency (JAXA), University of Bremen (Bremen), and Goddard Bootstrap (Goddard). Units are millions of square kilometers.
\end{figure}

In Figures  \ref{ice_dts_month} ($\lambda_S  {=}  1$) and \ref{ice_dts_month_seqnseq} ($\lambda_G  {=}  1$) we show optimal latent sea ice extent extractions ($\widehat{SIE^*}$) in black, together with the four raw indicators in color, by month. First consider Figure  \ref{ice_dts_month}.  Of course $\widehat{SIE^*}(\lambda_S {=} 1)$ is centered on $SIE^S$ due to the $\lambda_S  {=}  1$ normalization.  Moreover,  $\widehat{SIE^*}(\lambda_S {=} 1)$ is always very close -- almost identical -- to $SIE^S$ (and close to $SIE^B$, because $SIE^B$ tends to be very close to $SIE^S$).\footnote{Indeed when making Figure  \ref{ice_dts_month} we added a tiny constant (0.1) to $SIE^S$ to make it easier to distinguish $SIE^S$ from $\widehat{SIE^*}(\lambda_S {=} 1)$.}

Now consider  Figure \ref{ice_dts_month_seqnseq} ($\lambda_G {=} 1$). Due to the different normalization, $\widehat{SIE^*}(\lambda_G {=} 1)$ is centered not on $SIE^S$ but rather on $SIE^G$, so  $\widehat{SIE^*}(\lambda_G {=} 1)$ is shifted upward relative to $\widehat{SIE^*}(\lambda_S {=} 1)$.  The location of  $\widehat{SIE^*}(\lambda_G {=} 1)$ relative to  $SIE^G$, moreover,  clearly varies by month.  In winter months, it tends to be greater than  $SIE^G$, whereas in summer months it tends to be less than  $SIE^G$.  Note in particular that  the variation of $\widehat{SIE^*}(\lambda_G {=} 1)$ around $SIE^G$ is noticeably greater than the variation of $\widehat{SIE^*}(\lambda_S {=} 1)$ around $SIE^S$.  Clearly  $\widehat{SIE^*}(\lambda_G {=} 1)$ is influenced more by movements in other indicators ($SIE^S$, $SIE^J$, and $SIE^B$) than is $\widehat{SIE^*}(\lambda_S {=} 1)$.  

\begin{figure}[tbp] 
   	\caption{Extracted Sea Ice Extent, by Month}
   	\begin{center}
   		\includegraphics[scale=.25]{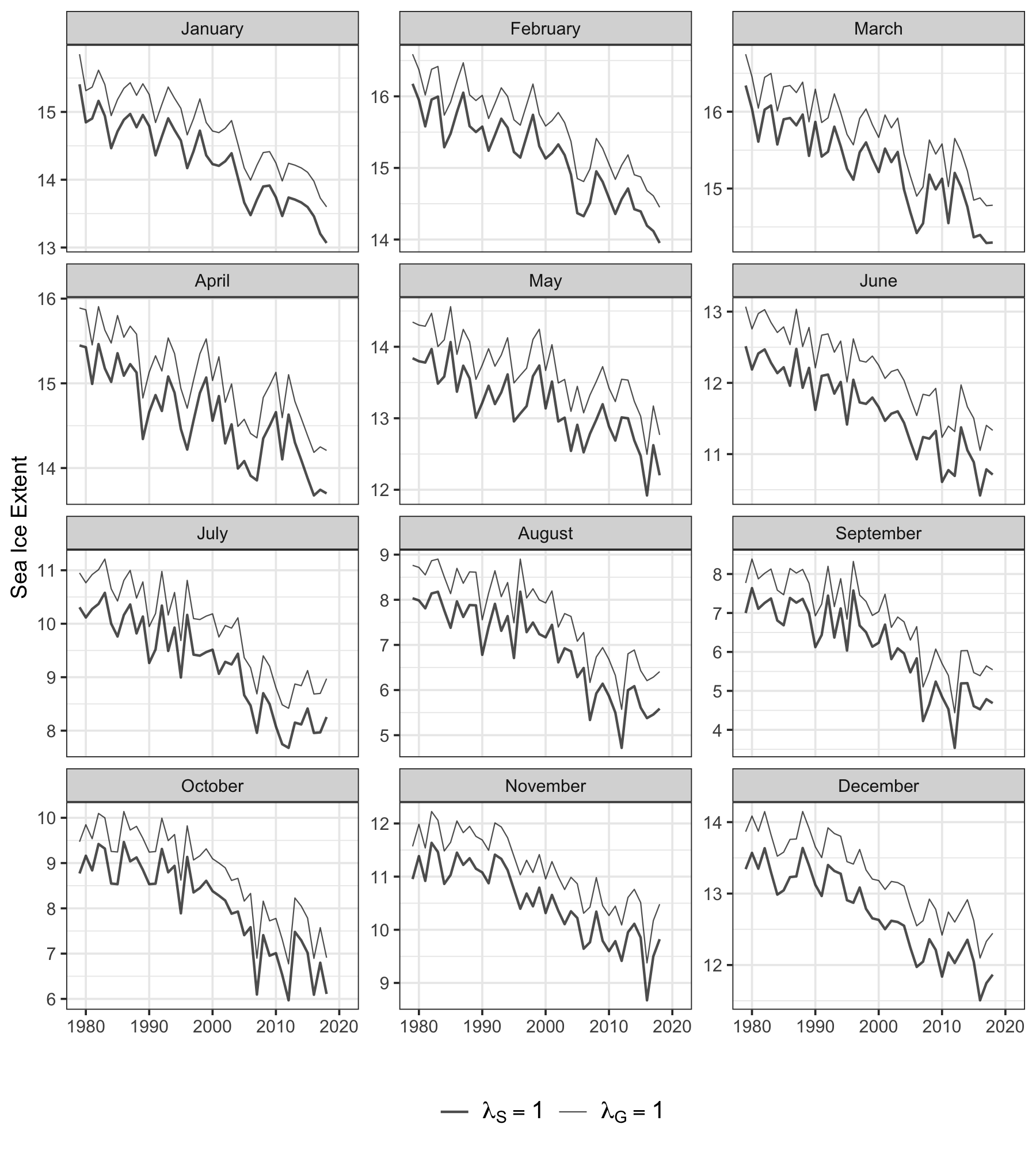}
   	\end{center}
   	\label{ice_dts_month_SvsG99}
   	Notes: Units are millions of square kilometers.
   \end{figure}

The fact that  $\widehat{SIE^*}(\lambda_S {=} 1)$ and $\widehat{SIE^*}(\lambda_G {=} 1)$ are  different, both in terms of  level  and variation around the level,  limits their usefulness for research focusing on level, as the level depends entirely on identifying assumptions.  However, and crucially, in an important sense  $\widehat{SIE^*}(\lambda_S {=} 1)$ and $\widehat{SIE^*}(\lambda_G {=} 1)$ are highly similar:  \textit{The model and identification scheme makes 
 	$\widehat{SIE^*}(\lambda_S {=} 1)$ and $\widehat{SIE^*}(\lambda_G {=} 1)$ identical up to a linear transformation.}  This is clear in Figure \ref{ice_dts_month_SvsG99}, which plots the two competing extracted factors, month-by-month. Regressions of $\widehat{SIE^*}(\lambda_G {=} 1)$ on $\widehat{SIE^*}(\lambda_S {=} 1)$ yield highly-significant intercepts not far from zero, 
highly-significant slopes near 1.0, and $R^2$ values above 0.999, for each month. 

Because  $\widehat{SIE^*}(\lambda_S {=} 1)$ and $\widehat{SIE^*}(\lambda_G {=} 1)$ are identical up to a linear transformation, it makes no difference which $\widehat{SIE^*}$ we use  for research focused on linear \textit{relationships} between $SIE^*$ and other aspects of climate (e.g., various radiative forcings).  The obvious choice, then, is $\widehat{SIE^*}(\lambda_S {=} 1)$, which is just $SIE^S$ itself, dispensing with the need to estimate the factor model.

%

\section{Summary and Conclusion}  \label{concl}

We propose a dynamic factor model for four leading Arctic sea ice extent indicators.  We estimate the model and use it in conjunction with the Kalman smoother  to produce a  statistically-optimal combination of the individual indicators, effectively ``averaging out" the individual measurement errors. We explore two identification strategies corresponding to two different factor loading normalizations. The corresponding two extracted combined measures (latent factors) are identical up to a linear transformation, so either one can be used to explore relationships of sea ice extent and other variables.  Interestingly, however, the extracted factor for one of the normalizations puts all weight on the Sea Ice Index.  Hence the Sea Ice Index alone is a statistically optimal ``combination" and one can simply use it alone with no loss, dispensing with the need to estimate the factor model.  That is,  there is no gain from combining the Sea Ice Index with other indicators, confirming and enhancing confidence in the Sea Ice Index and the NASA Team algorithm on which it is based, and similarly lending credibility -- in a competition against  very sophisticated opponents -- to the NSIDC's claim that the Sea Ice Index is the ``final authoritative SMMR, SSM/I, and SSMIS passive microwave sea ice concentration record" (\cite{NSIDC_SII}).  

\clearpage

\bibliographystyle{Diebold}
\addcontentsline{toc}{section}{References}
\bibliography{Bibliography}

\end{document}